\documentclass{article}

\usepackage[final]{neurips_2025}

\usepackage[utf8]{inputenc} 
\usepackage[T1]{fontenc}    
\usepackage{hyperref}       
\usepackage{url}            
\usepackage{booktabs}       
\usepackage{amsfonts}       
\usepackage{nicefrac}       
\usepackage{microtype}      
\usepackage{xcolor}         

\usepackage{booktabs}
\usepackage{amsmath}  
\usepackage{amssymb}  
\usepackage{threeparttable}  
\usepackage{graphicx}  

\title{The Influence of Establishing Belt and Road Node Cities on the Development of Digital Inclusive Finance Across Chinese Provinces}

\author{%
  Bo Wu \\
  Beijing International Studies University\\
  \texttt{2021221257@stu.bisu.edu.cn} \\
}

\begin{document}

\maketitle

\begin{abstract}
\vspace{-.5em}
The development of digital inclusive finance has become the trend of global inclusive finance development and an important strategy for China's national development. With the power of the Internet, big data and other digital technologies, digital inclusive finance has overcome the exclusivity and regionalism of traditional financial services, and can more efficiently meet the financial service needs of all types of people. "The Belt and Road" is the most important development strategy in China in recent years, in 2023, The State Council issued the "Digital China Construction overall Layout Plan" as the extension and expansion of traditional financial services, the development of digital inclusive finance has become an important strategic goal of the "Belt and Road" initiative.

\textbf{In 2016}, in order to promote the "New Silk Road Economic Belt" to drive the development of digital economy in provinces along the route, China proposed the concept of
"Digital Silk Road", and in the same year, the capital cities of ten provinces along the "New Silk Road Economic Belt", such as Henan, Shaanxi and Chongqing, were set up as node cities of the "Belt and Road" and "Digital Silk Road", giving full play to the driving role of provincial central cities. As an important part of digital finance and digital economy, digital financial inclusion has become one of the main development indicators of the Belt and Road Initiative. In October 2018, the "Belt and Road" Financial Inclusion International Forum was held in Lanzhou, a node city of the "Belt and Road". At the forum, scholars from various countries had profound exchanges on the development of digital financial inclusion and the opportunities and challenges faced by node cities.

\textbf{Domestic and foreign scholars} have conducted many studies on the influencing factors of digital inclusive finance, but few studies on the impact of the "Belt and Road" and the establishment of node cities on digital inclusive finance. Therefore, it is necessary to study the impact of the establishment of node cities in the "Belt and Road" on the development of digital inclusive finance. This paper mainly uses descriptive analysis and literature analysis to summarize and analyze the development status of China's digital inclusive finance and relevant theoretical research on the development of the "Belt and Road" Initiative, and analyzes and collates the background knowledge needed for the demonstration from the aspects of the development status of China's digital inclusive finance and the impact of digital inclusive finance on the economy. In this process, referring to the relevant economic theories, the theoretical model of this paper is proposed and the influence machine analysis is carried out.

\textbf{Empirically}, this paper selects the development level of digital inclusive finance in 31 provinces in China from 2011 to 2020 as the explained variable, takes the establishment of "Belt and Road" node cities as a quasi-natural experiment, and verifies the promoting effect of the establishment of "Belt and Road" node cities on the development of digital inclusive finance in provinces through the differential differential method. And verify whether the level of Internet development is a mediating variable. The empirical results show that the establishment of node cities in the "Belt and Road" does promote the development of digital inclusive finance in provinces with the level of Internet development as an intermediary variable.

key words: \textbf{The Belt and Road Initiative; Digital economy; Digital financial inclusion; Double difference method}
\vspace{-.5em}

\end{abstract}
\section{Introduction}
\subsection{Research Background and Significance}
\subsubsection{Research Background}
The concept of the digital economy was first proposed in 1996 by American economist Don Tapscott. Although he did not provide a clear definition at the time, he boldly predicted that the digital economy would become one of the most important components of 21st-century economic life. Following the implementation of the U.S. “Information Superhighway” initiative in the late 20th century, the concept of the digital economy was officially introduced into national governance. Entering the 21st century, countries around the world adopted active strategies to promote the digital economy, which began to flourish globally—validating Tapscott’s foresight.

Although the digital economy was introduced relatively late in China, it has experienced rapid and transformative growth. The Chinese government has actively invested in digital infrastructure and promoted interprovincial technological and industrial cooperation through the Belt and Road Initiative (BRI), significantly accelerating the digital economy's development. In particular, China has demonstrated outstanding performance in fields such as short video platforms and social media. With the release of the \textit{14th Five-Year Plan for Digital Economy Development}, digital technologies have been further elevated as a strategic focus of national economic development. As China approaches the threshold of the fourth industrial revolution, key challenges include enhancing the level and inclusiveness of digital economic development, advancing digital trade, and promoting industrial digitalization.\cite{su15108311}

In 2006, China’s “No.1 Central Document” introduced several policies targeting the emerging field of online lending, marking the start of growing domestic attention toward digital inclusive finance. As the global economy continues to evolve, traditional financial services increasingly struggle to meet public demand. Compared with traditional finance, digital inclusive finance offers broader coverage and faster services. In 2016, the State Council issued a national plan for the development of digital inclusive finance through 2020, signaling that China’s digital financial system had entered a mature phase. When the COVID-19 pandemic broke out in 2020, China’s digital inclusive finance system played a key role in helping small and micro enterprises withstand the crisis and resume production. According to Peking University’s Digital Inclusive Finance Index, from 2020 to 2023, the index grew by more than 5\% annually for three consecutive years. Since its inception, digital inclusive finance has played an increasingly vital role in China's economic life and has made a substantial contribution to the operation and growth of the digital economy and financial system.

As digital inclusive finance continues to evolve, new priorities have emerged, including cloud-based payments, big-data credit systems, online marketing, and cybersecurity. To further enhance development, China has issued a series of plans and measures, integrating digital technology with the Belt and Road Initiative to drive digital inclusive finance in the provinces along the route. Developing digital inclusive finance within the BRI framework offers the following advantages:

\begin{enumerate}
    \item Traditional economic strategies have often favored coastal and eastern provinces, leading to imbalances between coastal and inland areas. The “New Silk Road Economic Belt” encompasses ten inland provinces in northwestern, southwestern, and central China. By designating the capitals of these provinces as node cities, the government aims to leverage their influence to boost the digital economy and inclusive finance in inland regions, promoting balanced development between coastal and inland cities.
    
    \item The BRI covers a broad geographic range with many participating provinces, some of which are sparsely populated and geographically remote, with underdeveloped traditional financial services. Using these BRI provinces as nodes for developing digital inclusive finance maximizes its inclusiveness, allowing digital technologies to extend financial services to a broader population—especially to rural residents and remote regions.
\end{enumerate}

\subsubsection{Research Significance}

\paragraph{(1) Theoretical Significance.}  
This paper summarizes and analyzes existing research, and contributes to the theoretical literature on the policy factors affecting digital inclusive finance. By using the establishment of Belt and Road (BRI) node cities as a quasi-natural experiment, it focuses on the differences in provincial digital finance development before and after the intervention. Moreover, the paper integrates various theoretical frameworks such as China's central city development theories across different periods, the theory of digital inclusive finance, growth pole theory, siphon effect theory, and radiation effect theory. It ultimately confirms that the establishment of BRI node cities has played a significant role in promoting digital inclusive finance, providing theoretical support for future policy formulation.

\paragraph{(2) Practical Significance.}  
China’s digital inclusive finance has developed rapidly. While much research exists on its influencing factors and the role of the BRI in advancing the digital economy, there is limited literature exploring how the BRI affects provincial digital inclusive finance from a policy perspective. In addition, the financial service impact of the BRI remains insufficiently understood. Therefore, this study contributes to a more comprehensive understanding of the BRI’s influence mechanism on provincial digital inclusive finance, offering reference value for policymaking in China’s digital economy and digital trade development.\cite{hirose2008three}

\subsection{Research Content and Methodology}

\subsubsection{Research Content}

This thesis is divided into six main chapters:

\begin{itemize}
    \item \textbf{Chapter 1: Introduction.} This chapter introduces the research background and significance, outlines the research content and methodology, and highlights the innovations and limitations of the study.
    
    \item \textbf{Chapter 2: Literature Review.} This chapter reviews relevant theories on digital economy and digital inclusive finance, and summarizes key literature on the Belt and Road Initiative—particularly focusing on its impact on the digital economy and inclusive finance.
    
    \item \textbf{Chapter 3: Background Overview.} This chapter explains the contextual foundation of the study from three perspectives: the development of China’s digital economy, the development of digital inclusive finance, and the impact of the BRI on provincial digital financial inclusion.
    
    \item \textbf{Chapter 4: Theoretical Model and Mechanism Analysis.} This chapter combines relevant economic theories with the research topic, explaining how macro- and microeconomic theory guides the study. It further analyzes the influence mechanism and proposes theoretical hypotheses.
    
    \item \textbf{Chapter 5: Empirical Analysis.} The study treats the establishment of BRI node cities as a quasi-natural experiment, using non-node provinces as the control group and node provinces as the treatment group to empirically test the hypothesis that node cities enhance the development of digital inclusive finance. The entropy weight method is used to measure the level of internet development in each province, which is then tested as a mediating variable in the process.
    
    \item \textbf{Chapter 6: Policy Recommendations.} Based on the characteristics of China’s digital economy and digital inclusive finance, this chapter provides targeted recommendations for improving the level of financial inclusion, enterprise digitalization, and digital environment security.
\end{itemize}

\subsubsection{Research Methodology}

\begin{enumerate}
    \item \textbf{Literature Review Method.} The study begins by reviewing key concepts related to digital economy and digital inclusive finance, summarizing policy documents on the BRI and digital finance development, and thoroughly examining relevant macro- and microeconomic theories to build the theoretical model and conduct mechanism analysis.

    \item \textbf{Descriptive Analysis Method.} The study collects data on China's digital economy, digital inclusive finance, and the Belt and Road Initiative, and conducts descriptive analysis of the background information necessary for the research.

    \item \textbf{Empirical Analysis Method.} By treating the establishment of BRI node cities as a quasi-natural experiment, the study empirically tests their impact on provincial digital inclusive finance. It also uses the entropy weight method to assess provincial internet development levels and examines the mediating effect of this variable in the policy impact pathway.
\end{enumerate}

\subsection{Innovations and Limitations}

This paper employs a difference-in-differences (DID) approach to investigate the impact of the establishment of Belt and Road Initiative (BRI) node cities on the development of provincial digital inclusive finance in China. Additionally, it examines the mediating effect of internet development, representing an innovative research perspective.

While there exists a substantial body of literature evaluating the influencing factors of digital inclusive finance, relatively few studies integrate multiple dimensions—such as the BRI policy effect, internet development level, education level, and investment in high-tech industries—into a comprehensive assessment. By focusing on how the BRI affects the development of digital inclusive finance across provinces, this study offers findings with valuable policy implications, contributing an innovation in research content.\cite{fan2022research}

However, during the empirical analysis, it was observed that the mediating effect results for certain years were not robust. This may be due to the influence of other policies or the presence of unobserved moderating variables during those years. These issues merit further exploration and refinement in future research.
\section{Literature Review}
\subsection{Research on the Digital Economy}

The concept of the digital economy was first introduced in 1995 by American scholar Don Tapscott. Since then, scholars and research institutions worldwide have actively studied and expanded upon this concept. In 2000, the U.S. Department of Commerce identified e-commerce as a major direction for the development of the digital economy in its report \textit{Digital Economy 2000}. As the global digital trade network gradually took shape, the concept of the digital economy was embraced by countries around the world, and the tide of economic globalization further promoted the development of digital economy and digital trade. At the 2016 G20 Hangzhou Summit, China proposed the \textit{G20 Digital Economy Development and Cooperation Initiative}, which sparked global interest in the definition, classification, and measurement of the digital economy.\cite{cournede2015finance}

In 2018, at the International Conference of Official Statistics, experts Ahmad and Ribarsky proposed a new perspective in their report \textit{A Framework for Measuring the Digital Economy}. They suggested that data should be classified as a distinct type of product beyond physical goods, and that the communications industry—whose end products are data and information—should also be included within the scope of the digital economy.

Research on the development of China’s digital economy is abundant, particularly in regard to the effectiveness of national policies and issues related to international digital trade. Meng Qi (2023), through empirical analysis, clarified China's position in the global supply chain and identified weak links in the development of its digital economy. Wang Juan (2022) emphasized the importance of systematically summarizing the main paths of national digital economic development, scientifically measuring policy impacts, and analyzing policy factors affecting the digital economy, all of which are now frontier issues in academic circles. Zhang Wenkui (2022) argued that the development of the digital economy can promote industrial upgrading, while Zhan Xiaoning (2018) suggested that it helps improve the efficiency of foreign investment utilization.

\subsection{Research on the Development of Digital Inclusive Finance}

Currently, domestic research on digital inclusive finance is relatively extensive. As of September 2023, CNKI contains 3,349 articles using the keyword “digital inclusive finance,” and 25 articles using “Belt and Road + digital inclusive finance.” Yan Deli (2018) argued that U.S. digital economic policies have exacerbated the uneven development between the Global North and South. Zhou Luyao (2022) reviewed the development process of digital inclusive finance and qualitatively analyzed its economic impact. Yixin (2015) explained its influence on industrial structure. Zong Min (2019) studied the role of the Belt and Road Initiative in promoting digital inclusive finance in countries along the route. Yi Xingjian (2018) empirically demonstrated the impact of digital inclusive finance on household consumption. Wu Jinwang (2018) summarized and analyzed the influencing factors of digital inclusive finance.\cite{helms2006access}

These studies reveal that scholars’ research on digital inclusive finance mainly focuses on three aspects: its economic impact, influencing factors, and the effect of the Belt and Road Initiative on digital inclusive finance in participating countries. Digital inclusive finance has become an increasingly popular research topic; however, there is limited empirical literature examining how national strategies or policies influence digital inclusive finance. Rarely have studies investigated the impact of strategic plans or initiatives on the development of digital inclusive finance at the provincial level in China.

This paper takes the establishment of Belt and Road node cities as a quasi-natural experiment and conducts an empirical analysis of its impact on the development of provincial digital inclusive finance. The results confirm that the establishment of these node cities has significantly promoted the development of digital inclusive finance in China. The findings carry strong practical implications and provide policy suggestions for future development from the perspective of national strategy.

\subsection{Research on the Belt and Road Initiative and Node Cities/Provinces}

The concept of digital trade emerged relatively early, and after 2010, scholars from various countries began to show strong interest in it. For example, Chinese scholar Xiong Li was the first to provide a systematic overview and summary of digital trade on the international stage. However, the definition of digital trade remained vague for a long time. In 2013, the United States International Trade Commission (USITC), in its report \textit{Digital Trade in the U.S. and Global Economies}, formally defined digital trade as a trade model—both domestic and international—that relies on wired or wireless digital networks for the transmission of resources and information. In 2014, the USITC further expanded the scope of this concept to include trade conducted via digital platforms and cross-border information services.

In China, research on digital trade policies has grown along with the advancement of the Belt and Road Initiative (BRI). The impact of the “Digital Silk Road” on China’s digital economy and trade has become a major research focus. As of June 2022, CNKI lists 1,303 papers using the keywords “China + digital economy policy,” among which 707 discuss the Belt and Road Initiative and node cities, accounting for approximately 55\% of the literature on China's digital economy policy.

Proposed by President Xi Jinping in 2013, the Belt and Road Initiative has steadily developed over the past decade. The Digital Silk Road is an important extension of the BRI into the realm of digital economy, aiming to promote China’s domestic digital economic development and facilitate technological exchange and business cooperation with countries along the route, through the use of the Internet and related communication technologies.\cite{hasan2022promoting}

The idea of digitizing the Silk Road was first introduced by Professor Anupam Chander of the University of California, Davis. In his 2013 book \textit{The Electronic Silk Road: How the Web Binds the World Together in Commerce}, Chander emphasized that modern technology and information systems have expanded the channels of communication along the Silk Road. He proposed that with node cities acting as hubs, interactions that were historically impossible can now be realized.

Around the same time, China also began feasibility studies on the Digital Silk Road. In 2015, Zhu Yunqiang’s research team proposed a comprehensive plan to build the Digital Silk Road Economic Belt, under the national science and technology project \textit{Support Plan for Economic Belt and Information Infrastructure Construction of the Digital Silk Road}. Based on the current development trends of China’s digital economy, they argued that promoting the construction of the Digital Silk Road could significantly improve the level of digital economic development in the provinces along the route and enhance the efficiency of digital financial services through richer information resources.

In 2022, Zhang Jinlin’s team asserted that the development of digital inclusive finance contributes to achieving common prosperity, and that the establishment of Belt and Road node cities helps bridge the urban-rural digital divide and enhances the inclusiveness of the digital economy.

\subsection{Literature Commentary}

The above literature review reveals that existing studies have extensively examined the current state and influencing factors of digital inclusive finance development in China. The prevailing view is that the development of digital inclusive finance can enhance regional market vitality, promote the rationalization of industrial structure, and reduce the urban-rural income gap. For example, Huang Qian and her team (2019) analyzed the poverty reduction effects of digital inclusive finance and provided a detailed explanation of its connection with rural development. Li Zhiguo (2021) argued that the advancement of digital inclusive finance can effectively facilitate the transformation and upgrading of provincial industrial structures.

A broader look at domestic research on digital economy strategies and planning shows that the impact of the Belt and Road Initiative (BRI) and the Digital Silk Road on the development of the digital economy has become a prominent research focus in recent years. As mentioned earlier, Zhang Jinlin’s team (2022) emphasized that developing digital inclusive finance can promote common prosperity and that the Digital Silk Road may offer new opportunities for its advancement. Li Li (2017) suggested that the BRI could improve the efficiency of small and medium-sized enterprises (SMEs) along the route and enhance financial inclusiveness.\cite{yu2022digital}

Based on the above literature, this paper proposes the following research approach: the BRI and the Digital Silk Road are two of China's most significant recent initiatives. Their advancement is expected to promote regional welfare and facilitate the development of digital inclusive finance in provinces along the route. However, to what extent do these initiatives actually produce such effects? Does the establishment of Belt and Road node cities influence the development of digital inclusive finance at the provincial level? Are there mediating variables involved? These are the core empirical questions this study seeks to answer.

\section{Development of China's Digital Economy and Digital Inclusive Finance}
The digital economy serves as a crucial pillar of the modern industrial system. In 2022, the scale of China’s digital economy reached 50.2 trillion yuan, ranking second globally and marking a nominal year-on-year growth of 10.3\%. This was the second-highest growth rate in history, following only 2021. The share of the digital economy in GDP continued to rise, accounting for 41.5\% of the total. As an extension of traditional finance, digital inclusive finance has come to occupy an increasingly significant position within the digital economy.

This chapter focuses on summarizing the development history and current status of China's digital economy and digital inclusive finance. It also elaborates on the context of the Belt and Road Initiative (BRI) as a theoretical and empirical backdrop. These discussions are intended to enhance readers’ understanding of the development level of digital inclusive finance in China, the progress of the BRI, and the state of internet infrastructure.

\subsection{Development of China’s Digital Economy}
China's digital economy began relatively late. In September 2016, President Xi Jinping explicitly called for the comprehensive development of the digital economy during the G20 Hangzhou Summit. Prior to this, China had explored digital economic development in several areas—for example, the establishment of the “Digital Silk Road” and the issuance of the \textit{Outline for the Development of Big Data}. However, these efforts had not yet entered a phase of comprehensive development.

In 2017, at the G20 Hangzhou Summit, China led the launch of the \textit{G20 Initiative on Digital Economy Development and Cooperation}, marking the country's growing competitiveness and influence in the domains of digital economy and digital trade. Subsequently, the \textit{Guiding Opinions of the CPC Central Committee and the State Council on Promoting High-Quality Development of Trade} and the \textit{14th Five-Year Plan for Digital Economy Development}, both released in 2021, emphasized that from 2021 to 2025, China would enter a stage of accelerated digital economic development. During this period, the share of digital economy-related industries in GDP would continue to increase, and ongoing technological progress would not only drive the transformation of traditional industries but also lead to broader application of digital products in daily life. China's digital products have steadily expanded their global market share, showing consistent upward momentum.\cite{li2022digital}

In this new historical phase, China's digital economy carries the mission of promoting common prosperity. Enhancing the inclusiveness of digital services and narrowing the “digital divide” among provinces and between urban and rural areas have become critical issues in building socialism with Chinese characteristics. As a driving force and an engine of innovation on the path to common prosperity, the digital economy has significantly impacted both people's lifestyles and social production patterns. Technological progress has facilitated industrial structural adjustment and innovation, making the digital economy an increasingly essential component of China’s development strategy.

Moreover, the intrinsic inclusiveness of the digital economy contributes to the sharing of achievements across sectors such as the economy and people's livelihoods. In October 2022, the 20th National Congress of the Communist Party of China proposed higher requirements for digital economic development. On the industrial front, China should continue to advance the digital transformation of traditional industries and promote the widespread adoption of digital identities among enterprises to improve information transparency. On the administrative front, efforts should be made to improve the national integrated digital government service platform and expand public-oriented digital services.

Currently, China's digital economy strategy is twofold: on one hand, leveraging digital technologies to stimulate industrial development; on the other, embedding digital economy into daily life by promoting digitalization and shared services from the perspective of the general public.

In 2023, China had built and put into operation more than 2.3 million 5G base stations, with over 560 million 5G users nationwide—accounting for more than 60\% of the global total. China has thus become the country with the highest 5G network coverage and the largest number of gigabit cities in the world. Digital infrastructure has become the cornerstone of China’s efforts to enhance its industrial digitalization capabilities and to drive comprehensive digital transformation.

To balance digital economic development between the eastern and western regions, the Chinese government introduced the \textit{“Eastern Data and Western Computing”} (\textit{Dongshu Xisuan}) policy. This initiative aims to build an advanced and comprehensive nationwide computing infrastructure system and establish a large number of new physical data centers. It provides the eastern provinces with enhanced computing resources while leveraging networked computing systems to rebalance computational capacity between the eastern and western regions. The goal is to promote more equitable digital economic development and foster mutual learning in digital technologies. On one hand, the policy has improved the level of digitalization in transportation and logistics in western regions; on the other hand, it has supplied the computation-intensive eastern regions with valuable complementary resources.

In terms of big data and related infrastructure, China has achieved significant results in internet penetration and industrial digitalization. Following the release of the \textit{Action Plan for Promoting Big Data Development} by the State Council in 2016, the country has continued to improve upon the foundational infrastructure for big data as outlined in the plan. In 2017, the report of the 19th National Congress of the Communist Party of China called for the robust development of big data in the current stage and emphasized the need for comprehensive deepening and broadening of its application.

The implementation of the Belt and Road Initiative (BRI) has integrated remote provinces and cities into the national big data network. In 2021, the \textit{14th Five-Year Plan for the Development of the Big Data Industry} was issued, proposing the expansion of network coverage in four key industries, including chemical materials and logistics, and promoting industry-wide digitalization. That same year, the government approved information infrastructure projects across multiple provinces and townships to support the rapid growth of the digital economy. These efforts have further advanced the goal of achieving full broadband and signal base station coverage, thereby improving infrastructure, narrowing the urban-rural digital divide, and maximizing the institutional advantages of China’s development model.

\subsection{Development of Digital Inclusive Finance in China}
\subsubsection{Characteristics and Advantages of Digital Inclusive Finance}
Digital finance is an extension and evolution of traditional finance through the integration of digital technologies. As the market continues to expand, the demand for financial services from both residents and business operators has been increasing. Due to geographical constraints, traditional financial services are typically centered around physical financial outlets, which are often located in populous provincial economic centers. This setup has led to two major drawbacks:

\begin{enumerate}
    \item Financial outlets are concentrated in economically developed central areas of provinces. However, due to high population density, these outlets often suffer from inefficiencies, resulting in long queues and waiting times.
    \item Residents in peri-urban areas and relatively remote rural regions are unable to access financial services promptly due to their distance from such outlets.
\end{enumerate}

As the mismatch between the supply and demand of financial services intensifies, the traditional financial service system increasingly fails to meet the needs for borrowing, investment, savings, and other financial services. In response, digital inclusive finance has emerged as a more efficient and cost-effective alternative that is less constrained by geographical limitations.\cite{hasan2021does}

As a component of the broader financial system, inclusive finance is designed to serve groups that are typically underserved by traditional financial institutions—particularly those living in remote areas or with limited mobility. Unlike the exclusivity often associated with traditional financial services, inclusive finance is inherently accessible and is thus also referred to as “financial inclusion.”

\subsubsection{Policies on the Development of Digital Inclusive Finance in China}
This study organizes the development of digital inclusive finance in China by region for the year 2020 (data source: Peking University Digital Finance Research Center). The results indicate that digital inclusive finance is most developed in the eastern region, followed by the central region, and then the western region. To balance the regional development of digital inclusive finance, the State Council issued the \textit{Plan for Advancing the Development of Inclusive Finance (2016--2020)} in 2016. The plan emphasizes that the development of digital inclusive finance in China must adhere to the fundamental principles of equal opportunity and regional balance. It not only aims to improve the efficiency of digital financial services in provincial economic centers but also seeks to enhance inclusiveness, providing low-cost and demand-oriented financial products for residents in the eastern, central, and western regions. 

To meet the financial needs of the people, China has implemented numerous new policies and initiatives to comprehensively promote digital inclusive finance from multiple perspectives. The primary constraints on the development of digital inclusive finance are technological bottlenecks and environmental challenges. Therefore, this section focuses on two aspects of China’s policy framework: technological advancement and cybersecurity.

\paragraph{(1) Technological Advancement.} 
Since the issuance of the above-mentioned plan, China has continuously enhanced the development of core digital technologies, including the Internet, artificial intelligence, the Internet of Things, and blockchain. Following the release of documents such as the \textit{13th Five-Year Plan for Digital Rural Development}, the future direction for comprehensive digitalization in agriculture was clearly outlined. In the industrial and manufacturing sectors, the digital transformation is also progressing steadily. In 2022, the numerical control rate of key processes and the digitalization rate of machinery in industrial enterprises reached 58.6\% and 77.0\%, respectively. The total output value of digitalized industrial production exceeded 1.17 trillion yuan, representing an increase of 193 billion yuan over 2021. 

5G technology has been deeply integrated into industrial manufacturing. By the end of 2022, there were over 4,000 national projects under the “5G + Industrial Internet” initiative, and digitalized manufacturing applications accounted for 40.5\% of the entire manufacturing sector—signaling that China’s industrial sector has officially entered the digital age. In the financial and service industries, China has also undertaken extensive exploration and promoted digital transformation. Under the guidance of the central government’s \textit{Opinions on Several Policy Measures to Accelerate the Development of the Service Industry}, provinces have issued directives and subsidies for the development of digital finance and the service sector. For example, Hunan Province introduced the \textit{Several Policies on Promoting High-Quality Development of Modern Services}, ensuring the healthy growth of the digital tertiary sector.\cite{motzkin1953double}

\paragraph{(2) Cybersecurity.} 
China has actively introduced and refined laws and regulations related to the digital economy and digital inclusive finance. The continuous improvement and elaboration of the \textit{Cybersecurity Law} and the \textit{Critical Information Infrastructure Protection Law} have enabled enterprises in relevant regions to engage in trade more securely and legally, alleviating concerns about legal loopholes and promoting the rule-of-law foundation for digital economic and trade development.

Regarding critical information infrastructure security, the State Council issued the \textit{Regulation on the Security Protection of Critical Information Infrastructure} in 2021, which clearly defines the scope of critical infrastructure. Information security is no longer confined to inter-enterprise or inter-government concerns but also applies to data involving trade secrets, national security, and core technologies, which must receive enhanced protection. Safeguarding critical information is not only vital for national security but also for the security of enterprises and markets. 

According to American scholar Boles Ning in \textit{Market Environment and Trade Efficiency}, the market security index is significantly positively correlated with the average technological advancement and development scale of local enterprises. Therefore, ensuring the security of critical infrastructure is the cornerstone of a safe digital environment and information ecosystem, and it plays a pivotal role in supporting continuous digitalization and the advancement of digital technologies.

\begin{table}[htbp]
\centering
\caption{\textbf{Digital Financial Inclusion Development Index in China by Region (2020)}}
\begin{tabular}{lclclc}
\toprule
\textbf{Eastern Region} & \textbf{DFI} & \textbf{Central Region} & \textbf{DFI} & \textbf{Western Region} & \textbf{DFI} \\
\midrule
Zhejiang     & 406.88 & Hubei       & 358.64 & Chongqing   & 344.76 \\
Jiangsu      & 381.61 & Anhui       & 350.16 & Shaanxi     & 342.04 \\
Fujian       & 380.13 & Henan       & 340.81 & Sichuan     & 334.82 \\
Guangdong    & 379.53 & Jiangxi     & 340.61 & Guangxi     & 325.17 \\
Shandong     & 347.81 & Hunan       & 332.03 & Yunnan      & 318.48 \\
Hainan       & 344.05 & Shanxi      & 325.73 & Tibet       & 310.53 \\
Liaoning     & 326.29 & Inner Mongolia & 309.39 & Ningxia      & 310.02 \\
Hebei        & 322.70 & Jilin       & 308.26 & Xinjiang    & 308.35 \\
            &        &             &        & Guizhou     & 307.94 \\
\bottomrule
\end{tabular}
\label{1}
\\[1ex]
\footnotesize{Source: Peking University Digital Finance Research Center}
\end{table}
\subsubsection{Mechanism Analysis of the Impact of Digital Financial Inclusion on Regional Economies}
China has been promoting the development of digital financial inclusion networks by enhancing regional internet infrastructure, improving cybersecurity, and advancing digital technologies. With the rapid national development of digital financial inclusion, scholars have conducted in-depth studies on its driving factors. Current mainstream perspectives suggest that digital financial inclusion can foster regional economic development by promoting economic growth, narrowing urban-rural income gaps, and optimizing industrial structures.\textbf{Table~\ref{1}}

\paragraph{(1) Digital financial inclusion drives regional economic growth.} 
According to the “financial development theory” proposed by Swedish economist Knut Wicksell, under conditions of sound market credit, monetary actions and financial services exert a stimulating effect on economic activity. Unlike the classical monetary neutrality theory, this view suggests that when interest rates deviate from the natural rate and price levels are not constant, monetary and financial behaviors can positively influence market economies. Building upon Wicksell’s theory, economist Joseph Schumpeter emphasized that services provided by the banking system are essential drivers of economic development. A well-developed banking system facilitates the selection of qualified entrepreneurs—those with sound credit and potential are more likely to access funding—thereby fostering a healthier credit ecosystem through a “survival of the fittest” mechanism. As an extension of traditional finance, digital financial inclusion can stimulate regional economic vitality more broadly. It expands financial services into remote areas with underdeveloped transportation and enables more precise screening of entrepreneurs through big data technologies. Furthermore, cloud computing allows for accurate demand forecasting, thereby promoting the healthy development of local enterprises.\cite{carlino1997intercalation}

\paragraph{(2) Digital financial inclusion promotes industrial structure optimization.} 
With the emergence of the digital economy and continuous market liberalization, financial capital has increasingly shaped China’s investment landscape. In contrast to fiscal-driven investment models, financial capital has greater influence over industrial structure, affecting production scale, operational models, and the allocation of production factors. The sound development of digital financial inclusion can efficiently meet enterprises’ financial service needs, encouraging a shift toward technology-intensive industries, optimizing factor input structures, and ultimately enhancing both enterprise-level resource allocation and overall industrial structure.

\paragraph{(3) Digital financial inclusion helps bridge the urban-rural gap.} 
Reducing urban-rural disparities is widely recognized as one of the key functions of digital financial inclusion. Chinese scholar Tian Jie analyzed the relationship between digital financial inclusion and urban-rural disparity through the lens of financial exclusion. The theory of financial exclusion posits that certain social groups are marginalized from the financial system. These individuals may be physically disabled, lack literacy, or live in high-crime areas where financial institutions are scarce. As financial markets mature and institutions perform more granular segmentation, although such behavior may enhance credit systems and bring economic benefits, it also has drawbacks—groups in remote, impoverished, or less mobile conditions may be excluded. The inclusive nature of digital financial services can integrate these populations into the financial system. With access to the internet, individuals in remote or disadvantaged areas can obtain financial services locally, helping to meet the financial needs of rural residents, reduce the urban-rural income gap, and partly bridge the digital divide.

\subsubsection{Mechanisms Through Which Digital Inclusive Finance Influences Industrial Development}
Today, many countries around the world have embraced the strategy of optimizing industrial structures and revitalizing traditional industries through the advancement of digital inclusive finance. China, leveraging its strong research capabilities in the digital domain and extensive digital network infrastructure, stands at the forefront of this global trend. In 2022, the total output of China's digital economy exceeded 50.2 trillion yuan, ranking second globally. Compared to 2021, it contributed an additional 5.04 trillion yuan in output, further solidifying its role in the economy and accounting for 41.1\% of the total economic volume. With improvements in digital technologies and the gradual maturation of the digital inclusive finance system, China is steadily advancing the optimization of resource allocation and industrial structure. Digital inclusive finance has helped stimulate the development of traditional industries while also empowering digital economic growth with the foundational strength of China’s industrial capacity.

From the perspective of industrial sectors, digital inclusive finance is playing an increasingly important role in agriculture. China has made substantial progress in extending digital infrastructure and inclusive financial services to rural areas. On this basis, agricultural digitalization has begun to take shape. As of now, the digitalization rate of agricultural production nationwide has reached 24.9\%. High-efficiency digital agriculture technologies such as smart irrigation systems, precision temperature control in greenhouses, and intelligent fertilizer mixing—distinct from traditional farming methods—have gradually gained prominence. The release of the \textit{13th Five-Year Plan for the Development of Agricultural and Rural Informatization} has provided a clear roadmap for the comprehensive digital transformation of agriculture in the future.

In the manufacturing sector, the development of digital inclusive finance has become a key instrument for promoting industrial digitalization and optimizing the allocation of production factors. China has continuously introduced financing services specifically targeting small and medium-sized enterprises (SMEs). The relatively mature digital inclusive financial system has enabled rapid recovery of digital industries following the COVID-19 pandemic. By the end of 2022, the total market value of China’s digitalized industrial production exceeded 1.17 trillion yuan, representing a 193 billion yuan increase over 2021. 5G technology has been deeply integrated into industrial manufacturing. As of the end of 2022, over 4,000 national projects had been launched under the “5G + Industrial Internet” initiative, and digital industrial applications accounted for 40.5\% of the entire manufacturing sector. Digital inclusive finance has made a significant contribution to China's post-pandemic industrial recovery.

In the financial and service sectors, China has engaged in extensive exploration, striving to improve the quality of digital financial services through technological innovation. Under the guidance of the central government’s \textit{Opinions on Several Policy Measures to Accelerate the Development of the Service Industry}, provincial governments have issued directives and subsidy programs to promote the development of digital finance and service industries. For instance, Hunan Province implemented the \textit{Several Policies on Promoting the High-Quality Development of the Modern Service Industry}, ensuring the healthy growth of the digital tertiary sector.

From a regional development perspective, the Yangtze River Delta (YRD) urban agglomeration has demonstrated the most significant achievements in digitalization. In September 2021, Zhejiang Province reported that the total scale of the digital economy in the YRD region accounted for 43.3\% of the region's total income, and more than one-quarter of the national digital economy. In terms of digital economic inclusiveness, the YRD region places strong emphasis on the digital development of small and medium-sized enterprises (SMEs) and provides subsidies to support them. The region's industrial digitalization coverage has reached 71.3\%. The central government has encouraged and supported delegations from various provinces to study the successful practices of inclusive finance in the YRD region, with the aim of enhancing interregional cooperation in the digital economy, promoting positive interactions across regions and industries, and strengthening in-depth development.

From the perspective of enterprise financial service demand, in February 2023 the central government issued the \textit{Implementation Opinions on the High-Quality Development of Digital Inclusive Finance}, which proposed “lowering the difficulty of direct financing for enterprises and improving the efficiency of digital inclusive financial services.” Comprehensively improving digital financial services for micro, small, and medium-sized enterprises (MSMEs), and promoting not only key sectors but also inclusive development, has become a major institutional and policy advantage in China’s digital finance strategy.

At present, China has more than 47 million MSMEs and over 61.5 million individual small merchants. MSMEs form the backbone of the Chinese economy, accounting for 91.5\% of the total market, providing 82\% of employment, and contributing over 63\% of gross national income. In the current context of sluggish global trade and the lingering impact of the pandemic, many MSMEs—especially those dependent on import and export—have faced significant challenges. To stabilize the “fundamentals” of enterprise development, China has implemented a series of supportive policies for MSMEs since 2021. These efforts aim to enhance the coverage of digital inclusive finance, ensuring that policies truly benefit every MSME, while continuously promoting innovation and expansion in inclusive finance through both financial and technological support.

As the level of digital inclusive finance improves, banks and financial institutions can more efficiently evaluate and assess the operating status of MSMEs through digital networks and computational technologies, enabling them to offer targeted financial support and loan solutions based on local conditions.

\subsection{The Impact of Establishing Belt and Road Node Cities on Provincial Digital Economy Development}
International cooperation in the digital domain is a key mechanism for promoting high-level opening-up. During the implementation of the Belt and Road Initiative (BRI), China has actively participated in discussions and consultations on digital issues under various international organizations and multilateral mechanisms, contributing to global digital governance from the Chinese perspective. China has also expanded the “Silk Road E-commerce” initiative, vigorously promoted digital trade, and continuously deepened international exchange and cooperation in the digital domain.

The “Digital Silk Road” is the digital extension of the Belt and Road Initiative (BRI), focusing on cooperation in cutting-edge fields such as the digital economy and artificial intelligence. It leverages digital and internet technologies to form a 21st-century Digital Silk Road. The establishment of the Digital Silk Road aims to foster digital economic integration and collaboration among BRI provinces and partner countries, raise the development level of participating regions, balance disparities in digital infrastructure construction, and promote the exchange of digital technologies across provinces.

As a critical component of the BRI, the Digital Silk Road and the development of the digital economy along its route are mutually reinforcing. From a cost perspective, the advancement of internet technologies can effectively promote coordinated economic development among provinces along the BRI corridor by lowering regional development costs and improving inter-provincial connectivity and collaboration. From the perspective of information efficiency, the Digital Silk Road enhances the inclusiveness of digital finance along its route. The more widespread digital inclusive financial services are, the lower the degree of information asymmetry. Commercial banks can assess clients’ creditworthiness more efficiently, while consumers can better understand the operational status of companies behind their consumption choices.

The establishment of BRI node cities has significantly promoted the development of digital inclusive finance in the provinces along the BRI route. This inclusiveness and balance are key characteristics distinguishing China's digital economic development model from that of the United States and other Western countries. The growth of the digital economy and internet technologies has removed geographical barriers to financial services. Digital inclusive finance eliminates reliance on physical branches, greatly improving service efficiency.

The development of digital inclusive finance has led to increased lending efficiency in rural areas and improved operational efficiency for financial institutions. In terms of financial institution operations, big data and artificial intelligence reduce operational costs, enabling traditional financial institutions to expand services using saved resources. Closely linked to network development, digital inclusive finance leverages the power of big data dissemination to minimize information asymmetry and imbalance. This enables more efficient identification of individuals in need of credit and other financial services, effectively lowering access thresholds.

Generally speaking, digital inclusive finance is characterized by relatively high fixed costs but low variable costs and wide coverage, making it well-suited to promoting coordinated urban-rural development and narrowing the income gap between urban and rural areas.\cite{waldhauser2001hypodd}

The Belt and Road Initiative and the development of digital inclusive finance are mutually reinforcing. The establishment of BRI node cities stimulates the growth of digital finance in those provinces, while improved digital financial services, in turn, enhance the financing efficiency of local digital enterprises and support technological advancement in internet-related fields. Given the timeliness and scalability of the internet, digital inclusive financial services offer more efficient information exchange and lower communication costs than traditional financial services.

One of the key advantages of the digital financial network is its relatively low long-term development cost. Most investment in building a digital inclusive finance system occurs in the initial phase. Once internet infrastructure is established, only minimal maintenance costs are required. According to Moore’s Law, as the scale of the internet expands, the average return increases while costs decrease. Once an information flow is established, the associated returns continue to rise, and the labor cost of maintenance becomes negligible in comparison.

Another core advantage of digital financial services is their timeliness. Information flows efficiently through fiber-optic and broadband networks, allowing banks and other financial institutions to promptly identify and respond to the needs of enterprise clients.

\section{Theoretical Framework and Mechanism Analysis}
This chapter first introduces several key macro- and microeconomic theories relevant to the present study and uses them to explain the main arguments of the paper. It then analyzes the mechanisms through which the establishment of Belt and Road (B\&R) node cities may influence the development of digital inclusive finance. Finally, it proposes a hypothesis and predicts potential empirical results, thereby providing a theoretical foundation for the empirical analysis in Chapter 5.

\subsection{Theoretical Foundations for the Impact of Belt and Road Node Cities on Digital Inclusive Finance}
\subsubsection{Regional Division of Labor Theory}
Adam Smith proposed that the division of labor originates from free trade. Whether within the same industry, the same region, or across the global market, continued development of free trade ultimately leads to more precise and scientific patterns of internal specialization. This division can be based on natural endowments or on advantageous production conditions formed through policy incentives, living costs, or other factors. Once a division of labor is established, countries, regions, or industries benefit from improved production efficiency and more effective resource utilization. If all actors adhere strictly to the principles of specialization, it will lead to the creation of maximum material wealth.

Building upon Smith’s ideas, economist Bertil Ohlin introduced the theory of regional division of labor. According to Ohlin, regional division is a relatively stable structure. Each region has its own natural endowments and production advantages, which shape its particular mode of specialization. Despite regional differences, the division of labor generally follows several common principles:

\begin{enumerate}
    \item Once regional specialization matures, regions tend to focus on producing a limited number of goods for which they are best suited. However, the value of these goods is often realized through interregional trade, not solely within the producing region.
    
    \item The level of infrastructure development in a region is positively correlated with the degree of industry specialization.
    
    \item Regional specialization often leads to the emergence of core economic zones, which, when supported economically, can drive the development of industries across the broader region.
    
    \item All forms of regional specialization and the emergence of core economic zones are the natural outcomes of the pursuit of economic efficiency and require no external intervention.
\end{enumerate}

According to the theory of regional division of labor, when the establishment of the Belt and Road Initiative and the new “Digital Silk Road Economic Belt” node cities was proposed in 2016, the digital economies within provinces had already formed relatively mature patterns of specialization. The designation of node cities supports the development of regional core economic zones, thereby promoting the growth of the overall digital economy in those regions.\cite{zambaux1998influence}

\subsubsection{Growth Pole Theory}
The Growth Pole Theory was proposed by French economist François Perroux. The theory posits that as economies and markets continue to develop, leading enterprises within an industry and firms with strong research capabilities tend to cluster in core economic zones within a region. As capital becomes increasingly concentrated, economies of scale are achieved, thereby reducing operational costs across all enterprises and sectors while improving production efficiency. These core zones are referred to as “growth poles,” which can exert economic radiation effects on surrounding areas. By spreading cost and R\&D advantages from the growth pole outward, the development of relevant industries is stimulated, resulting in more balanced development between core and peripheral areas within the region.

The impact of a growth pole on other parts of the region mainly manifests in two aspects:

\begin{enumerate}
    \item Through the radiation effect, resources and technologies from the growth pole diffuse throughout the region. A portion of these returns to the growth pole due to free trade dynamics, forming a cyclical flow. Once a growth pole is established, market forces alone can sustain its continued development.

    \item Due to the “leakage effect,” when national or provincial governments provide subsidies and preferential policies to growth pole areas, enterprises in these areas can drive the development of similar industries across the region through technological progress and innovation. Since growth poles usually contain a large number of medium- and large-sized enterprises, their advancement enhances regional consumption capacity and stimulates market vitality.
\end{enumerate}

The designation of Belt and Road (B\&R) node cities aligns well with the Growth Pole Theory. On one hand, China promotes the development of digital economy and digital inclusive finance in node cities, which in turn drives the overall regional digital development through the radiation effect. On the other hand, comprehensive policy support for non-node cities within node provinces—including investments in digital infrastructure, industrial digitalization, and digital security—has elevated digital economic growth in underdeveloped areas (including remote, rural, and border regions). The resulting return flows and agglomeration further strengthen node cities, creating a virtuous cycle of provincial development.

\subsubsection{The Long Tail Economy Theory}
The Long Tail economy theory was proposed by Chris Anderson, who analyzed product sales across markets and identified a normal distribution pattern in product sales. The “long tail” refers to the large number of products with relatively low sales, which together form the long tail of the distribution. These products typically exhibit two key characteristics: (1) although each market segment is relatively small, the segmentation is fine-grained; and (2) while each niche market is small in scale, their aggregate number is substantial. 

In the context of internet markets, as long as smooth supply and distribution channels are in place—along with adequate storage and transportation conditions—these “niche markets” can achieve market shares comparable to, or even exceeding, those of popular mainstream products.

These small markets within the internet domain are referred to as “niche markets.” According to Philip Kotler, niche markets possess significant potential. Realizing this potential depends on whether the markets’ demands for storage, logistics, marketing, and distribution can be effectively met. If a national or regional government can concentrate efforts to support these niche markets with the necessary infrastructure and material resources, they can yield substantial returns in both production and services.

According to the Long Tail theory, the development of the Belt and Road (B\&R) Initiative and its designated node cities has provided the material foundation for the growth of niche markets within provinces. The construction of transportation infrastructure has reduced logistics costs for small markets, while the rapid development of node cities has created rising demand for the production of non-mainstream products. This has led to the proliferation of various e-commerce platforms and small-goods trading websites.\cite{zhang2006development}

Take, for example, the trade of rural small commodities. Such transactions emphasize cost-effectiveness and show low brand dependency. The advancement of the B\&R and the Digital Silk Road has accelerated the development of digital networks in both urban and rural areas. Products that once required travel to weekly markets can now be purchased via local WeChat groups or small-goods e-commerce platforms from neighboring villages or counties. Coupled with increasingly efficient transportation systems, rural and urban consumers now favor online purchasing and home delivery services.

In summary, the B\&R Initiative has promoted the expansion of regional internet infrastructure, and its influence in driving the development of niche markets within the internet economy has been particularly significant.

\subsubsection{Theoretical Perspectives on the Determinants of Digital Inclusive Finance}
American economists Sarma and Pais~\cite{augustinis2012analise} were among the first to study the determinants of digital inclusive finance from a macroeconomic perspective. They found that a nation's level of modernization and Human Development Index (HDI) are positively correlated with the development of digital inclusive finance. Building on this, economist Anand~\cite{anand2007bioavailability} concluded that both regional GDP and the level of digital economy development are positively associated with digital inclusive finance.

Chinese scholar Xia Pingfan et al.~\cite{abbott2021gwtc} investigated the mechanisms through which the internet influences the development of digital inclusive finance and found a robust correlation between internet penetration and the advancement of digital financial inclusion. Currently, there are two mainstream views in academia regarding this mechanism:

\begin{enumerate}
    \item The increase in internet penetration enhances payment efficiency and diversifies payment methods, thereby fostering the development of internet finance. Since internet finance can effectively promote digital inclusive finance, higher internet penetration rates will boost the development of digital financial inclusion.
    
    \item The development of internet and big data technologies enhances the security of data sharing, enables broader data accessibility, breaks down “information silos,” and thereby promotes the development of digital inclusive finance.
\end{enumerate}

Based on existing research into the determinants of digital inclusive finance, this paper hypothesizes that the implementation of the Belt and Road (B\&R) Initiative has stimulated digital economic growth in node cities and their respective provinces by improving infrastructure and regional economic strength. Furthermore, drawing upon the “long tail theory,” the B\&R Initiative has expanded market access for non-mainstream products, encouraging the rapid development of online niche markets and ultimately accelerating the development of regional digital financial industries. Therefore, the B\&R Initiative may exert additional positive impacts on the development of digital inclusive finance.

\subsection{Mechanism Analysis}
In 2014, Premier Li Keqiang delivered a keynote speech titled “Jointly Creating a New Future for Asian Development” at the opening ceremony of the Boao Forum for Asia Annual Conference, in which he proposed new plans and adjustments to advance the Belt and Road Initiative. In 2016, to accelerate the development of the New Silk Road Economic Belt and leverage the driving force of participating provinces, the government divided the belt into three parts—northwest, southwest, and central—and designated ten provincial capitals as node cities.

According to the “growth pole theory” and the “regional specialization theory,” the designation of node cities has stimulated the digital economy in their respective provinces. As regional centers, these provinces further contribute to national digital economic growth. Moreover, drawing on the “long tail theory,” the B\&R Initiative has also facilitated internet development in the provinces where node cities are located. Based on the earlier analysis of the factors influencing digital economy development, regional economic development and internet penetration are positively correlated with digital inclusive finance. Thus, it is reasonable to assume that the establishment of B\&R node cities may have a significantly positive impact on the development of digital inclusive finance in their respective provinces. This paper proposes three primary mechanisms through which this impact may occur:

\subsubsection{Balancing the Permeation Effect and the Siphon Effect}
According to the growth pole theory, within a province, priority can be given to the development of central regions, which in turn can drive urban-rural development in non-central areas through the “permeation effect.” As production factors and technology from central regions continuously spread into surrounding areas, this promotes balanced regional development. The “siphon effect” serves as a sustainable guarantee of the central region's driving force. In a given region, since enterprises seek to minimize production and R\&D costs, firms and capital naturally concentrate in the central region under free market conditions. Therefore, as resources and technologies diffuse outward, the siphon effect can replenish the central region's economic growth, while the agglomeration of technological industries sustains continuous upgrades in production capabilities.

During the development of the Belt and Road Initiative (B\&R) and the Digital Silk Road, China designated “node cities” to fully leverage the leading role of central regions and to balance the development of the digital economy within provinces. According to theoretical models, after production factors circulate within a province, final goods often re-converge to central regions—typically the largest logistics hubs in the province—thus ensuring continued development of node cities. However, when market conditions fluctuate significantly, firms tend to adopt more conservative investment strategies, often favoring more mature market environments in central cities. In such scenarios, the siphon effect may surpass the permeation effect, leading to “over-concentration” of resources and technology in central regions.

To avoid this over-concentration, the Chinese government has adopted a more inclusive and balanced development approach when promoting node cities and their respective provinces:

\begin{enumerate}
    \item \textbf{Balancing digital infrastructure construction across provinces.} China has implemented a more balanced plan for building digital infrastructure. While constructing digital infrastructure along the B\&R Economic Belt, efforts have been made to ensure not only intra-provincial balance between node cities and remote urban-rural areas, but also national-level balance, as many node cities are located in the underdeveloped central and western provinces. The government has made significant efforts to accelerate digital infrastructure development in rural and remote areas.
    
    \item \textbf{Promoting digital inclusive finance in rural areas.} As the B\&R Initiative advances and node cities' financial service capabilities improve, the “digital divide” between urban and rural areas widens due to the siphon effect. The market alone struggles to address this disparity. Traditional financial institutions often find it difficult to serve remote rural areas due to geographic constraints, leaving large-scale agricultural producers such as crop farmers and livestock breeders without access to financial services. Therefore, it is challenging to extend digital inclusive finance to the countryside through market mechanisms alone. In response, the Chinese government has promoted internet infrastructure in remote villages and pioneered new models of rural inclusive finance, such as deploying technical personnel to rural areas and providing localized training. This initiative fosters local digital talent in townships and integrates rural communities into the digital society. With the widespread adoption of digital inclusive finance, a single computer terminal can enable farmers to obtain needed loans, improving both financial inclusion and the coordination of urban-rural development—thus achieving a dynamic balance between the siphon and permeation effects.
\end{enumerate}

\subsubsection{New Production Factors and Increasing Returns to Scale}
With the growing importance of the digital economy and digital trade in economic life, data has been recognized by modern economic theory as a new type of production factor. The 19\textsuperscript{th} National Congress of the Communist Party of China emphasized that, to deepen industrial reform, optimizing the market-based allocation of production factors must be a future priority. At this meeting, data was officially acknowledged as a production factor on par with traditional factors such as land, labor, and technology. Since then, the input structure in data-related industries has undergone significant transformation. The integration of this new key factor has reshaped the production factor system, becoming an essential complement to traditional factors and economic growth theory.

Unlike traditional production factors, data is replicable, shareable, and reusable, thereby overcoming the limitations of scarcity and exclusivity. These properties provide strong conditions for increasing returns to scale in related industries. In any given industry, the higher the proportion of data in total factor inputs and the deeper the integration between data and traditional factors, the greater the marginal returns of all inputs. This breaks through the growth rate limitations of endogenous growth theory in traditional sectors. During the production process, the inclusion of data as a production factor generates amplification, superposition, and multiplier effects on economic growth. When data is combined with traditional factors such as labor and technology, it significantly enhances the return on those inputs.

The establishment of Belt and Road (B\&R) node cities has driven the development of regional internet infrastructure and the digital economy, improving the efficiency of data utilization as a production factor. Through the efficient use of data resources, these regions achieve increasing returns to scale in both financial and digital industries, while simultaneously enhancing the inclusiveness of digital finance and overall production efficiency.

\subsubsection{Radiation Effect}
Regional central cities naturally emerge due to factors such as free trade and historical culture, and play a role in organizing and guiding the economic, political, and cultural life within a province. Generally, these cities are also the provincial capitals. According to the theory of regional radiation, the influence of regional central cities can extend throughout the entire province. Among various aspects of influence, including the economy, culture, technology, and politics, the economic radiation capability is the most dynamic.

Designating provincial capitals along the “Silk Road Economic Belt” as Belt and Road (B\&R) node cities can maximize the role of regional central cities in driving the digital economy. According to the “radiation effect theory”, this driving effect is mainly reflected in three aspects:

First, the advancement of the B\&R Initiative has promoted the development of infrastructure in provincial central regions. As infrastructure construction improves, transportation within the region becomes more efficient and warehousing costs are reduced. Transportation conditions and logistics costs are crucial factors influencing e-commerce development, and improvements in infrastructure levels also promote the development of provincial internet infrastructure. Additionally, transportation networks have incorporated more remote villages and towns into the main provincial transport system, reducing the cost of popularizing digital infrastructure (such as network base stations, IoT platforms, and big data centers) and enhancing the efficiency of inclusive financial services.\cite{monteiller2005efficient}

Second, the establishment of B\&R node cities has facilitated the exchange of digital technology within provinces. By designating regional central cities as B\&R node cities, the technological driving force of central cities can be fully leveraged. According to the “growth pole theory”, in order to minimize departmental costs and maximize economies of scale, high-tech and innovative enterprises tend to set up R\&D or main functional departments in node cities. Due to the radiation and technology diffusion effects, promoting digital technology development in node cities indirectly boosts the digital technology progress of the entire province, enhancing the efficiency of industrial digitalization.

Third, the establishment of B\&R node cities has improved the efficiency of digital inclusive financial services in provinces. Node cities serve as provincial financial centers, and enhancing financial service quality in node cities can effectively promote the development of inclusive digital finance across the province. Currently, the development of digital inclusive financial services targeting micro, small, and medium-sized enterprises (MSMEs) is one of the key focuses of China’s digital economy. There are over 47 million MSMEs and more than 61.5 million individual businesses in China, with MSMEs being the backbone of China’s economic development, accounting for 91.5\% of the total market. Developing provincial digital inclusive finance through the radiation effect of central cities is an important measure for stabilizing the “basic market” of enterprise development. As the digital financial service level in central cities continues to improve, the coverage rate of inclusive digital finance in provinces also increases. Provincial digital financial networks can efficiently utilize networked data and computational technology, enabling banks and financial institutions to better assess and account for the status of MSMEs and to provide tailored funding and loan support accordingly.

\begin{table}[htbp]
\centering
\caption{\textbf{Heterogeneity Analysis of DFI by Province (2020)}}
\resizebox{\textwidth}{!}{%
\begin{tabular}{lclrrrrr}
\toprule
\textbf{Province} & \textbf{Region} & \textbf{DFI2020} & \textbf{DFI Rank} & \textbf{Change since 2018} & \textbf{Change since 2015} & \textbf{Change since 2012} \\
\midrule
Hubei        & Central & 358.64 & 8  & -1 & 1  & 5  \\
Anhui        & Central & 350.16 & 9  & 1  & 8  & 9  \\
Henan        & Central & 340.81 & 14 & 1  & 11 & 10 \\
Jiangxi      & Central & 340.61 & 15 & -2 & 4  & 7  \\
Hunan        & Central & 332.03 & 17 & 2  & 5  & 2  \\
Shanxi       & Central & 325.73 & 19 & 2  & 4  & -2 \\
Inner Mongolia & Central & 309.39 & 25 & 4  & -9 & -2 \\
Jilin        & Central & 308.26 & 27 & -3 & -7 & -1 \\
Chongqing    & Western & 344.76 & 11 & 0  & 0  & -1 \\
Shaanxi      & Western & 342.04 & 13 & 1  & 0  & -2 \\
Sichuan      & Western & 334.82 & 16 & 0  & -2 & -4 \\
Guangxi      & Western & 325.17 & 20 & -2 & 1  & -5 \\
Yunnan       & Western & 318.48 & 22 & 4  & 4  & 3  \\
Tibet        & Western & 310.53 & 23 & 3  & 8  & 8  \\
Ningxia      & Western & 310.02 & 24 & 3  & -9 & -3 \\
Xinjiang     & Western & 308.35 & 26 & 2  & -2 & 1  \\
Guizhou      & Western & 307.94 & 28 & -5 & 2  & 1  \\
\bottomrule
\end{tabular}%
}
\label{2}
\\[1ex]
\footnotesize{Source: Peking University Digital Finance Research Center}
\end{table}
\begin{figure}[htpb]
	\centering
	\hspace{-1.5mm}
	\includegraphics[width=0.4\linewidth]{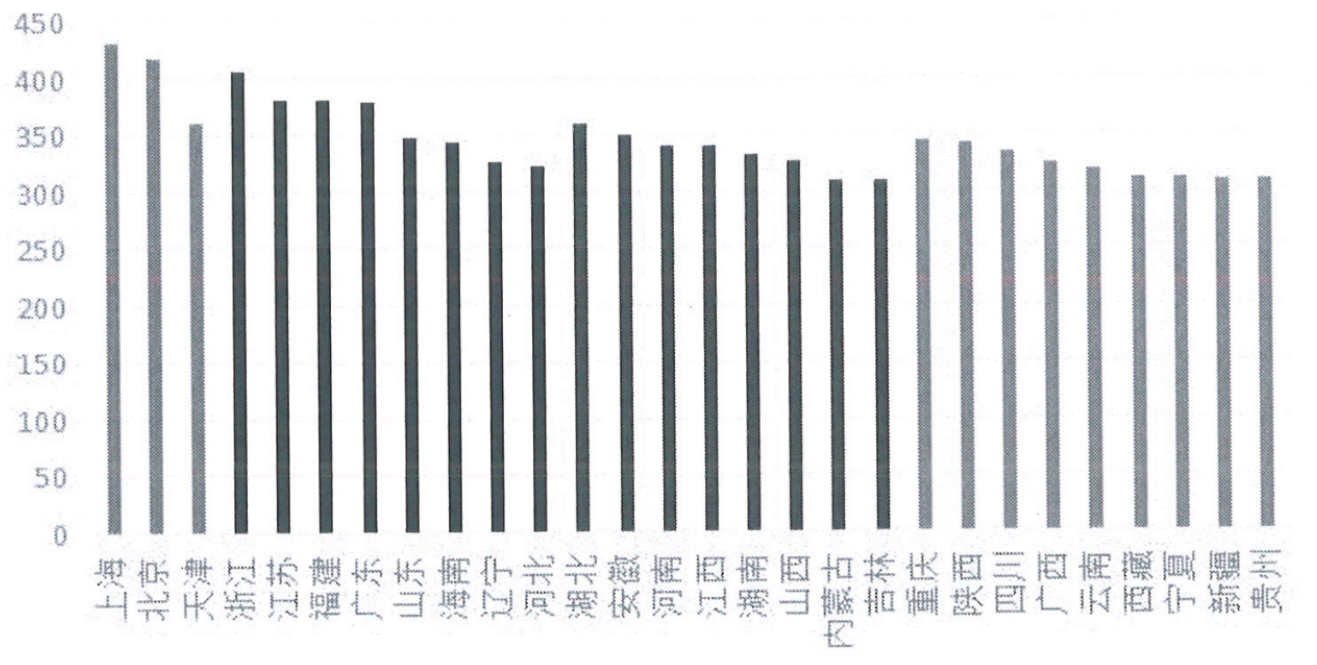}
	\caption{\textbf{China's Digital Financial Inclusion Development Index by Region and Province in 2020}}
	\vspace{-1.5mm}
	\label{f1}
\end{figure}
\subsection{Heterogeneity Analysis}
China has designated the capital cities of ten provinces in the central and western regions as key nodes for the “Belt and Road Initiative” (BRI) and the “New Silk Road Economic Belt.” Accordingly, this study compiles relevant data on the development of Digital Financial Inclusion (DFI) in central and western China from 2012 to 2020 (see \textbf{Table~\ref{2}}). Based on China’s regional classification, the central region includes eight provinces: Inner Mongolia, Shanxi, Hubei, Jilin, Anhui, Hunan, Heilongjiang, and Henan. The western region comprises eleven provinces and autonomous regions: Chongqing, Xinjiang, Sichuan, Tibet, Yunnan, Shaanxi, Guizhou, Guangxi, Ningxia, Gansu, and Qinghai.

According to the data presented in \textbf{Table~\ref{2}}, the development of DFI in the central region has generally outpaced that of the western region. This discrepancy may be attributed to two main factors. First, the eastern region, where DFI development began earlier and progressed rapidly (see \textbf{Figure~\ref{f1}}), has exerted a spillover effect on nearby areas. Given their closer geographical proximity to the east, central provinces have benefited more easily from the mature DFI systems of the eastern region. Second, the central region has a denser distribution of traditional financial service institutions. Provinces and cities with higher population density tend to have larger regional economic centers and more financial institutions. Thus, as the BRI advances simultaneously across provinces, those with a more developed traditional financial sector experience more significant improvements in DFI. Therefore, when evaluating the impact of BRI node cities on provincial DFI development, heterogeneity analysis should be duly considered.
\section{Empirical Analysis of Factors Influencing the Development of China’s Digital Economy}
\subsection{Theoretical Hypotheses and Sample Selection}
After a thorough review of relevant macroeconomic and microeconomic theories concerning node cities and the theoretical determinants of digital financial inclusion, this study establishes the mechanism through which the establishment of Belt and Road Initiative (BRI) node cities influences digital financial inclusion. As a result, the Difference-in-Differences (DID) method is chosen to analyze the impact of BRI node city establishment on provincial digital financial inclusion levels in China. The justification for using DID lies in the fact that the establishment of BRI node cities can be treated as a quasi-natural experiment, making DID an efficient tool for policy evaluation. Moreover, DID helps to mitigate the influence of endogeneity during the policy implementation process, thus enhancing the robustness of the empirical results.\cite{zhang2003double}

This paper empirically examines whether the establishment of “Digital Silk Road” node cities promotes provincial digital financial inclusion and contributes to the inclusiveness of the digital economy. Based on the theoretical framework and prior research, including Ge Heping and Zhu Huiwen (2018) and Xiao Yuanhao (2022), the following hypotheses are proposed:

\begin{itemize}
    \item \textbf{Hypothesis 1:} The establishment of BRI node cities positively promotes the development of digital financial inclusion at the provincial level.
    \item \textbf{Hypothesis 2:} The promotion effect of BRI node cities on digital financial inclusion is mediated by the improvement in internet development.
\end{itemize}

Given the central region's advantages in population density, geographical location, and infrastructure compared to the western region, and following the viewpoints of scholars such as Huang Qing (2019) and Zhou Luyao (2022), a third hypothesis is proposed:

\begin{itemize}
    \item \textbf{Hypothesis 3:} The effect of BRI node city establishment on digital financial inclusion exhibits heterogeneity, with a stronger impact observed in central provinces than in western provinces.
\end{itemize}

This paper treats the establishment of BRI node cities as a quasi-natural experiment. Provinces with BRI node cities are regarded as the treatment group, while those without such cities are treated as the control group. A Difference-in-Differences approach is used to assess the effect of node city establishment on the growth of digital financial inclusion.

To satisfy the preconditions of the DID method, the study uses panel data from 31 Chinese provinces over the period 2011–2020. The data source is the Digital Finance Research Center at Peking University, and the dataset includes the Digital Financial Inclusion (DFI) Index, which serves as the primary measure of digital financial inclusion. The DFI Index comprises 33 indicators, including payments, money market funds, credit, insurance, investment, and credit information. These indicators are standardized and weighted using methods such as the analytic hierarchy process to obtain the final index.

The treatment group includes provinces with established Digital Silk Road node cities: Henan, Shaanxi, Chongqing, Anhui, Hunan, Hubei, Qinghai, Gansu, Sichuan, and Jiangxi. The remaining provinces constitute the control group.

\subsection{Model Specification and Variable Definition}
\subsubsection{Difference-in-Differences Model Specification}
To examine the impact of establishing ``Belt and Road'' hub cities on the development of digital inclusive finance across provinces, this paper adopts a Difference-in-Differences (DID) approach. The following DID model is constructed to estimate the effect:

\begin{align}
Y_{it} = \beta_0 + \beta_1 did_{it} + \alpha X_{it} + \eta_i + \mu_t + \varepsilon_{it}
\label{e1}
\end{align}

Where $Y_{it}$ denotes the level of digital financial inclusion development (DFI), which is the core dependent variable. The variable $did_{it}$ is the interaction term for the DID estimation and is used to estimate the impact of establishing ``Digital Silk Road'' hub cities on provincial digital financial inclusion. In \textbf{Formula~\ref{e1}}, $i$ denotes province ($i = 1, 2, \ldots, 31$), $t$ denotes year ($t = 2011, 2012, \ldots, 2020$), and $Y_{it}$ is the DFI level of province $i$ in year $t$. 

The DID estimator $did_{it}$ is constructed as follows. The provinces are divided into treatment and control groups according to policy implementation. The variable \texttt{ydy1 = 1 \& after = 1} identifies the treatment group after the policy; in this case, $did_{it} = 1$, otherwise it is 0. Here, \texttt{after = 0} indicates the pre-policy period. If \texttt{ydy1 = 0} and \texttt{after = 1}, the province is in the control group after the policy. Conversely, \texttt{ydy1 = 1} and \texttt{after = 0} refers to the treatment group in the pre-policy period.

In 2013, President Xi Jinping proposed the ``Silk Road Economic Belt'' during his speeches in Kazakhstan and Indonesia. In 2016, the government announced the establishment of 10 pilot hub cities in regions such as Shaanxi, Qinghai, Gansu, Xinjiang, and Chongqing. We treat 2016 as the starting year for Digital Silk Road hub construction. The variable $X_{it}$ includes other control variables affecting digital financial inclusion.

\begin{table}[htbp]
\centering
\caption{\textbf{Evaluation Indicators, Direction of Impact, and Weights Based on the Entropy Method}}
\begin{tabular}{lccc}
\toprule
\textbf{Indicator} & \textbf{Direction of Impact} & \textbf{Weight} & \textbf{Data Source} \\
\midrule
Number of Domains                         & Positive & 0.0372 & China Statistical Yearbook \\
Number of Web Pages                       & Positive & 0.0138 & China Statistical Yearbook \\
Number of IPv4 Addresses (1st)           & Positive & 0.0228 & China Statistical Yearbook \\
Number of IPv4 Addresses (2nd)           & Positive & 0.0789 & China Statistical Yearbook \\
Broadband Internet Access Ports          & Positive & 0.1121 & China Statistical Yearbook \\
Mobile Internet Users                    & Positive & 0.4133 & China Statistical Yearbook \\
Mobile Internet Access Traffic           & Positive & 0.1404 & China Statistical Yearbook \\
Broadband Internet Access Users          & Positive & 0.0394 & China Statistical Yearbook \\
\bottomrule
\end{tabular}
\label{3}
\end{table}
\subsubsection{Mediation effect model setting}
In order to examine the mediating effect of the Internet development level LID (Level of Internet development) in promoting the development of inclusive digital finance in my country through the “Belt and Road” initiative, this paper refers to the mediating effect test method of Wen Zhonglin (2005) and establishes the following model to empirically test whether the Internet development level plays a mediating role in promoting the development of digital inclusive finance in provinces through the establishment of “Belt and Road” node cities. See \textbf{Formula~\ref{e2}}, \textbf{Formula~\ref{e3}} and \textbf{Formula~\ref{e4}} for details.

\begin{align}
Y_{it} = \theta_0 + \theta_1 did_{it} + \theta_2 y + \varepsilon_{it}
\label{e2}
\end{align}

\begin{align}
Z_{it} = C + \eta did_{it}
\label{e3}
\end{align}

\begin{align}
Y_{it} = C_1 + \lambda_1 did_{it} + \lambda_2 Z_{it}
\label{e4}
\end{align}

Here, $Y_{it}$ denotes the digital financial inclusion development level of province $i$ in year $t$, and $Z_{it}$ denotes the intermediary variable representing the level of internet development.

This paper adopts the entropy method to measure the level of Internet development. In view of the fact that the direct assignment method and the expert scoring method are relatively subjective, the entropy method can eliminate the interference of human factors as much as possible, and realize the weight assignment to measure the level of Internet development in China from 2011 to 2020 as much as possible. At the same time, the entropy method can also overcome the problem that the numerical differences of the selected economic indicators are too small, which makes analysis difficult, and thus can more scientifically mine the implicit information in the data. The core idea of this method is that the greater the degree of variation of the values of each indicator, the greater the weight should be assigned to it, and vice versa. The specific steps of its calculation are as follows: The first step is to use the range method to standardize the various indicators of the Internet development level from 2011 to 2020 to eliminate the influence of the dimensional factors of each indicator, and then obtain dimensionless indicators. All indicators are positive indicators, so the standardized \textbf{Formula~\ref{e5}} corresponding to the positive indicators is selected as follows:

\begin{align}
x'_{ij} = \frac{x_{ij} - \min(x_j)}{\max(x_j) - \min(x_j)}
\label{e5}
\end{align}

\textbf{Step 2:} Calculate the proportion of sub-indicator $j$ for province $i$ in year $t$: \textbf{Formula~\ref{e6}}

\begin{align}
Y_{ij} = \frac{x'_{ij}}{\sum x'_{ij}}
\label{e6}
\end{align}

\noindent
\textbf{Step 3:} Compute the information entropy of indicator $j$: \textbf{Formula~\ref{e7}}

\begin{align}
e_j = -k \sum (Y_{ij} \ln Y_{ij}), \quad \text{where } k = \frac{1}{\ln m}, \quad e_j \in [0, 1]
\label{e7}
\end{align}

\noindent
\textbf{Step 4:} Calculate the redundancy of information entropy using: \textbf{Formula~\ref{e8}}

\begin{align}
d_j = 1 - e_j
\label{e8}
\end{align}

\noindent
\textbf{Step 5:} Compute the weight of indicator $j$: \textbf{Formula~\ref{e9}}

\begin{align}
w_i = \frac{d_i}{\sum d_i}
\label{e9}
\end{align}

\noindent
\textbf{Step 6:} Compute the weighted sub-indicator value: \textbf{Formula~\ref{e10}}

\begin{align}
s_{ij} = w_i \times c_{ij}
\label{e10}
\end{align}

Summing over all sub-indicators yields the composite indicator for a given period.

The constructed Internet development index consists of domain name count, number of ports, IPv4 address count, and mobile Internet access traffic. Let $Y$ represent the level of Internet development, where $x_1$, $x_2$, $x_3$, and $x_4$ denote domain names, number of ports, IPv4 addresses, and mobile Internet traffic, respectively. The composite score is thus: \textbf{Formula~\ref{e11}}

\begin{align}
Y = x_1 + x_2 + x_3 + x_4 + \cdots + x_7
\label{e11}
\end{align}

From a structural perspective, the differences in Internet development levels across provinces and years originate from variation in these 7 core dimensions. The variance decomposition method can effectively capture how divergence in these dimensions drives Internet development disparities across regions and years.

According to the steps of entropy calculation mentioned above, the weights of Internet development indicators in various provinces in China from 2011 to 2020 can be calculated using Excel software, and the results are shown in \textbf{Table~\ref{3}}. The entropy method can achieve weight assignment more objectively to measure the Internet development status of each province as much as possible. The weight value implies and reflects the development characteristics of the research object. From the ranking of indicator weights in \textbf{Table~\ref{3}}, in the Internet development level evaluation system, the top two are the number of mobile Internet users and the number of Internet involved traffic, which shows that the number of mobile Internet users and the number of Internet involved traffic are important factors in determining the level of Internet development.

\subsubsection{Variable Selection and Descriptions}
\begin{enumerate}
  \item \textbf{Explained Variable (DFI).}  
  This paper investigates the impact of the establishment of “Belt and Road” node cities on the development of digital inclusive finance across provinces. The Digital Financial Inclusion (DFI) Index for each province per year, obtained from the Digital Finance Research Center at Peking University, serves as the most appropriate explained variable. The evaluation system of the DFI Index comprises 33 indicators covering payments, money market funds, credit, insurance, investment, and creditworthiness. The final index is calculated through weighted aggregation and reflects the annual development level of digital inclusive finance in each province.

  \item \textbf{Core Explanatory Variable \textit{treat} $\times$ \textit{post} (DID).}  
  This dummy variable is the product of two components: whether the province contains a “Belt and Road” node city, and whether the year is before or after the establishment of such a node. As the core of the Difference-in-Differences (DID) approach, this variable effectively mitigates endogeneity issues. In this study, provinces with node cities after their establishment year are coded as 1, and all others as 0.

  \item \textbf{Control Variables ($X_i$).}  
  These variables are introduced to control for various differences among cities, minimizing the impact of inter-city heterogeneity on the results. Following the studies of Ge Heping and Zhu Wen (2018), and Song Xiaoling (2017), the following control variables are selected:
  \begin{enumerate}
    \item \textit{Financial Awareness / Education Level (Edu)}: Financial awareness indicates a region's potential for digital inclusive finance. Due to its difficulty in quantification, this paper adopts Song Xiaoling’s (2017) proxy method—using higher education enrollment rate as a proxy for education level.
    \item \textit{Foreign Direct Investment (FDI)}: A major driving force for the development of digital inclusive finance at the regional level.
    \item \textit{Digital Enterprise Research Funding (DRF)}: Reflects regional capacity for digital research, which may influence digital inclusive finance.
    \item \textit{Disposable Personal Income per Capita (DPI)}: Indicates the penetration level of digital technology and is an important metric for assessing digital financial inclusiveness.
    \item \textit{Population Density (PD)}: Incorporates geographical and transportation conditions. Including population density as a control variable helps account for geographic factors (such as transportation and communication) affecting the cost of digital inclusive finance development.
  \end{enumerate}
\end{enumerate}

\begin{table}[htbp]
\centering
\caption{\textbf{Descriptive Statistics of Sample Variables}}
\begin{tabular}{lccrrrr}
\toprule
\textbf{Variable} & \textbf{Symbol} & \textbf{Observations} & \textbf{Mean} & \textbf{Std. Dev.} & \textbf{Min} & \textbf{Max} \\
\midrule
Explained Variable           & DFI   & 310 & 216.2   & 97.03        & 16.22     & 431.9     \\
Core Explanatory Variable    & DID   & 310 & 0.161   & 0.368        & 0         & 1         \\
Education Level              & Edu   & 310 & 82.95   & 40.70        & 6         & 167       \\
Foreign Direct Investment    & FDI   & 310 & 16085   & 25828        & 208       & 179268    \\
Research Expenditure         & DRF   & 310 & 916224  & 2.087e+06    & 196       & 2.348e+07 \\
Disposable Income            & DPI   & 310 & 27246   & 12499        & 9740      & 79610     \\
Population Density           & PD    & 310 & 46.54   & 73.62        & 0.264     & 414.8     \\
\bottomrule
\end{tabular}
\label{4}
\end{table}
\subsection{Descriptive Statistics}
An analysis of the descriptive statistics in \textbf{Table~\ref{4}} reveals that among the variables including the total Digital Financial Inclusion (DFI) Index, education level, foreign direct investment (FDI), and digital industry research funding (DRF), the greatest fluctuations are observed in FDI and DRF. This can be attributed to the fact that prior to the implementation of the national “Belt and Road” (B\&R) policy, some regions had limited economic cooperation with other B\&R countries. Following the policy’s introduction, provinces along the B\&R route increased collaboration with foreign enterprises in response to the national strategy, resulting in a sharp surge in FDI in a short period. Another reason for the dramatic change in FDI is that the B\&R Initiative expanded the scope of trade for Chinese provinces—especially for inland regions, whose trade potential was previously constrained by the lack of connectivity, which has now been improved through the business environment brought about by the overland Silk Road.

As for DRF, its variation can be explained by two main factors. First, before the B\&R and the “Digital Silk Road” initiatives, provinces had relatively low trade volumes and limited revenue. With the policy-driven income growth, investment in research has increased after meeting basic development needs. Second, enterprises, having achieved profitability, tend to invest more in research to move up the value chain and gain competitive product advantages.

From a regional perspective, the level of digital financial inclusion in inland provinces is significantly lower than that in coastal provinces. There are also notable disparities in FDI and research funding. However, since the introduction of the B\&R and Digital Silk Road initiatives, inland provinces have seen relatively rapid development in the digital economy driven by node cities along the overland Silk Road. Coastal provinces, benefiting from increased trade revenue and technological exchange via the Maritime Silk Road, have also accelerated their digital economy development. Nevertheless, the development gap between inland and coastal provinces has gradually narrowed. The descriptive statistics highlight the policy effect of rapid digital economic growth in provinces with node cities, although disparities in inherent endowments lead to considerable variation in development speed across provinces.

\subsection{Empirical Results and Heterogeneity Analysis}
\begin{table}[htbp]
\centering
\caption{\textbf{Regression Results}}
\begin{threeparttable}
\begin{tabular}{lcccccc}
\toprule
& (1) & (2) & (3) & (4) & (5) & (6) \\
& \multicolumn{6}{c}{\textbf{Dependent variable: DFI}} \\
\midrule
\textbf{did}           & 83.39***  & 78.56***  & 96.92***  & 69.07**   & 56.92***  & 62.79***  \\
                       & (14.24)   & (14.24)   & (14.15)   & (30.00)   & (9.642)   & (9.334)   \\
\textbf{Education}     &           & 0.328**   & -0.0304   & 0.143     & 0.00738   & -0.0440   \\
                       &           & (0.129)   & (0.142)   & (0.148)   & (0.0945)  & (0.0913)  \\
\textbf{FDI}           &           &           & 0.00115***& -0.000109 & -0.00296*** & 0.00145*** \\
                       &           &           & (0.000225)& (0.000408)& (0.000319)& (0.000425)\\
\textbf{DRF}           &           &           &           & 1.37e-05*** & 2.97e-05*** & 1.46e-05*** \\
                       &           &           &           & (5.10e-06)& (3.39e-06)& (4.40e-06)\\
\textbf{DPI}           &           &           &           &           & 0.00682*** & 0.00784*** \\
                       &           &           &           &           & (0.000343)& (0.000385)\\
\textbf{PD}            &           &           &           &           &           & -0.439*** \\
                       &           &           &           &           &           & (0.0856)  \\
\textbf{Constant}      & 202.8***  & 176.4***  & 184.6***  & 169.8***  & 41.08***  & 26.60**   \\
                       & (5.717)   & (11.83)   & (11.48)   & (5.717)   & (10.80)   & (10.76)   \\
\midrule
Year. Ind              & YES       & YES       & YES       & YES       & YES       & YES       \\
Observations           & 310       & 310       & 310       & 310       & 310       & 310       \\
R-squared              & 0.191     & 0.219     & 0.288     & 0.304     & 0.651     & 0.679     \\
\bottomrule
\end{tabular}
\begin{tablenotes}
\footnotesize
\item \textit{Note}: Standard errors in parentheses. \\
\item *** $p<0.01$, ** $p<0.05$, * $p<0.1$
\end{tablenotes}
\end{threeparttable}
\label{5}
\end{table}
\subsubsection{Empirical Results on the Impact of the Belt and Road Initiative on Digital Financial Inclusion}
Using Stata 11.0 for regression estimation, the coefficient of the interaction term \textit{did} (i.e., \textit{time} $\times$ \textit{treat}) captures the impact of the establishment of “Digital Silk Road” node cities on the development of digital financial inclusion. To prevent multicollinearity among control variables, a stepwise regression method is employed, gradually introducing the control variables. In \textbf{Table~\ref{5}}, Model (1) controls only for \textit{did}, along with city and year fixed effects. Model (6) presents the full specification, controlling for \textit{Education} (education level), \textit{FDI} (foreign direct investment), \textit{DRF} (enterprise R\&D expenditure), \textit{DPI} (disposable personal income per capita), and \textit{PD} (population density).

The coefficient on \textit{did} in Column (4) of the table is statistically significant at the 5\% level, while in all other columns, the coefficient is significant at the 1\% level. All coefficients are positive, indicating that the establishment of BRI node cities significantly promotes the development level of digital financial inclusion in the respective provinces.

As more control variables are added, the coefficient of the interaction term \textit{did} gradually decreases and stabilizes. When all control variables are included in Model (6), the estimated coefficient of \textit{did} is 62.79, confirming the positive impact of the establishment of BRI node cities on the development of digital financial inclusion. Therefore, the policy effect of “Digital Silk Road” node cities is both statistically and economically significant. Node cities have played a strong leading role as regional hubs, and after their designation, the digital financial inclusion level in the corresponding provinces improved significantly. Thus, Hypothesis 1 is supported.\cite{gerdan1995comparison}

The regression results also reveal several additional findings: Foreign direct investment (FDI) exerts a modest positive effect on digital financial inclusion in China. Digital enterprise research funding (DRF) is positively correlated with the level of digital financial inclusion. Disposable personal income per capita (DPI) also shows a positive correlation. However, population density (PD) is negatively correlated with digital financial inclusion, possibly because in some regions, digital infrastructure development has lagged behind rapid population growth. This may lead to a decline in digital financial inclusion despite economic vibrancy and active markets. In such areas, enterprise loan efficiency remains low, and the urban-rural digital divide is more pronounced.

\begin{table}[htbp]
\centering
\caption{\textbf{Mediation Effect Regression Results}}
\begin{threeparttable}
\begin{tabular}{lccc}
\toprule
& (1) & (2) & (3) \\
& \textbf{DFI} & \textbf{DFI} & \textbf{LID} \\
\midrule
\textbf{did}        & 77.12***  & 64.22***  & 0.0741***  \\
                    & (17.94)   & (14.61)   & (0.0264)   \\
\textbf{LID}        &           & 21.67**   &            \\
                    &           & (8.33)    &            \\
\textbf{Constant}   & 178.5***  & 193.7***  & 0.519***   \\
                    & (24.70)   & (26.02)   & (0.00830)  \\
\midrule
Year. Ind           & YES       & YES       & YES        \\
Observations        & 310       & 310       & 310        \\
R-squared           & 0.393     & 0.490     & 0.380      \\
\bottomrule
\end{tabular}
\begin{tablenotes}
\footnotesize
\item \textit{Note}: Standard errors in parentheses. \\
\item *** $p<0.01$, ** $p<0.05$, * $p<0.1$
\end{tablenotes}
\end{threeparttable}
\label{6}
\end{table}
\subsubsection{Test of Mediation Effect}
The test for mediation effects follows a three-step approach (Wen Zhonglin, 2022).

\textbf{Step 1:} Examine the impact of the core explanatory variable on the dependent variable.  
Models (1) and (2) in \textbf{Table~\ref{6}} report the effects of the explanatory variables on the dependent variable—digital financial inclusion development (DFI). Model (1) does not include the mediator variable LID (Level of Internet Development), while Model (2) presents the full regression results with the mediator included. According to Model (1), the core explanatory variable \textit{DID} has a significantly positive effect on the dependent variable, validating Equation (2).

\textbf{Step 2:} Examine the impact of the core explanatory variable on the mediator.  
Model (3) reports the effect of the core explanatory variable \textit{DID} on the mediator variable LID. The results show that \textit{DID} has a significantly positive effect on LID, validating Equation (3).

\textbf{Step 3:} Include the mediator variable in the regression and examine the effects of both the mediator and the core explanatory variable on the dependent variable.  
According to Model (2), once the mediator variable is included in the regression, the estimated coefficient of \textit{DID} on digital financial inclusion decreases from 77.12 to 64.22. This suggests that the level of internet development (LID) plays a partial mediating role in the process by which the establishment of BRI node cities promotes provincial digital financial inclusion. The estimated mediation effect is 12.9, which confirms Equation (4).

The establishment of BRI node cities promotes digital technology development in provinces by exerting the radiation effect of central cities. According to the "Long Tail Theory," the presence of node cities stimulates the growth of niche markets, fosters the development of the digital financial service sector, and ultimately enhances the level of internet development in provinces. Improved internet development directly enhances the efficiency of digital financial services and reduces service costs, thus contributing to the advancement of digital financial inclusion. Empirical evidence confirms that internet development level serves as a significant mediator in the process through which the establishment of BRI node cities promotes digital financial inclusion. The mediation effect accounts for 16.7\% of the total effect.

\begin{table}[htbp]
\centering
\caption{\textbf{Heterogeneity Test Results}}
\begin{threeparttable}
\begin{tabular}{lcccc}
\toprule
& (1) & (2) & (3) & (4) \\
& \textbf{DFI (M)} & \textbf{DFI (M)} & \textbf{DFI (R)} & \textbf{DFI (R)} \\
\midrule
\textbf{did}         & 118.8***  & 94.83***  & 103.3***  & 82.05***  \\
                     & (18.21)   & (27.90)   & (18.61)   & (22.83)   \\
\textbf{Education}   &           & -0.502    &           & 1.338**   \\
                     &           & (0.663)   &           & (0.574)   \\
\textbf{FDI}         &           & -0.0121** &           & -0.0184***\\
                     &           & (0.00556) &           & (0.00635) \\
\textbf{DRF}         &           & 0.000127*** &         & 8.31e-05** \\
                     &           & (3.47e-05)  &         & (3.96e-05) \\
\textbf{DPI}         &           & 31.08*     &         & 17.15      \\
                     &           & (39.78)    &         & (31.97)    \\
\textbf{PD}          &           & -0.293     &         & 0.0623     \\
                     &           & (0.967)    &         & (0.847)    \\
\textbf{Constant}    & 174.6***  & 263.2***   & 175.6*** & 153.5***   \\
                     & (9.597)   & (39.82)    & (8.874)  & (18.02)    \\
\midrule
Observations         & 90        & 90         & 110      & 110        \\
R-squared            & 0.326     & 0.471      & 0.222    & 0.499      \\
\bottomrule
\end{tabular}
\begin{tablenotes}
\footnotesize
\item \textit{Note}: Standard errors in parentheses. \\
\item *** $p<0.01$, ** $p<0.05$, * $p<0.1$
\end{tablenotes}
\end{threeparttable}
\label{7}
\end{table}
\subsubsection{Heterogeneity Test}
Based on the earlier analysis of heterogeneity between China’s central and western provinces, this study conducts separate DID regressions for the DFI of the central region (M) and the western region (R), with the results presented in \textbf{Table~\ref{7}}. The regression results clearly show that, both before and after the inclusion of control variables, the establishment of BRI node cities has had a significantly positive effect on digital financial inclusion in both central and western provinces—significant at the 1\% level. Moreover, the impact is stronger in central provinces compared to western provinces, thus validating Hypothesis (3).

\subsection{Robustness Tests}
\begin{figure}[htpb]
	\centering
	\hspace{-1.5mm}
	\includegraphics[width=0.4\linewidth]{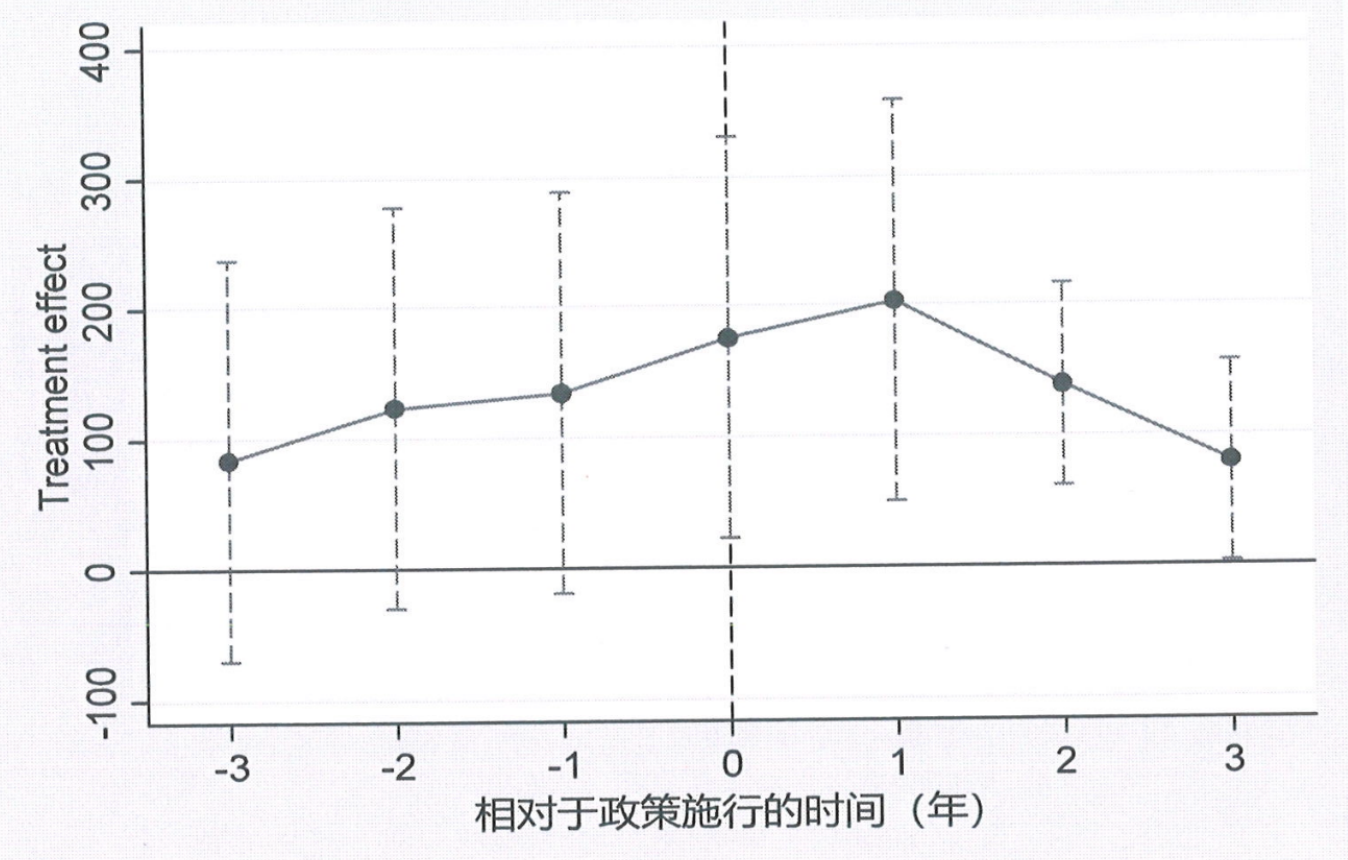}
	\caption{\textbf{Parallel trend test chart}}
	\vspace{-1.5mm}
	\label{f2}
\end{figure}
\subsubsection{Parallel Trend Test}
The use of the difference-in-differences (DID) method requires that the treatment and control groups exhibit similar trends prior to the policy intervention. This is because DID relies on the counterfactual logic that assumes the control group reflects the pure time effect of policy-unaffected development. If the treatment group had not been influenced by the policy, its time trend should mirror that of the control group. Therefore, it is essential to test the consistency of the digital financial inclusion development trends between the treatment and control groups over the period 2011–2020. \textbf{Figure~\ref{f2}} shows that during this time, both groups experienced similar overall trends in digital financial inclusion, thereby satisfying the parallel trend assumption.

\subsubsection{Placebo Test}
A placebo test is conducted by artificially shifting the policy implementation year to test the robustness of the results. If the estimated effects are no longer significant under the placebo year, this would confirm the robustness of the original findings. Here, we assume that the policy was implemented in 2018 and rerun the regression. As shown in Model (2) of Table 5-5, the interaction term \textit{treat} $\times$ \textit{year} is not statistically significant, and we fail to reject the null hypothesis. This supports the conclusion that the original positive effects of BRI node city establishment on digital financial inclusion are robust and not driven by spurious correlations.\cite{waldhauser2000double}

\begin{table}[htbp]
\centering
\caption{\textbf{Robustness Test Results}}
\begin{threeparttable}
\begin{tabular}{lcccc}
\toprule
& (1) & (2) & (3) & (4) \\
& \textbf{DFI'} & \textbf{DFI'} & \textbf{DFI'} & \textbf{DFI'} \\
\midrule
\textbf{did}         & 42.1      & 46.7      & 62.71**   & 69.07**   \\
                     & (21.05)   & (17.31)   & (31.04)   & (30.00)   \\
\textbf{Education}   &           & 0.139     &           & 0.143     \\
                     &           & (0.137)   &           & (0.148)   \\
\textbf{FDI}         &           & 0.000377  &           & -0.000109 \\
                     &           & (0.000386)&           & (0.000408)\\
\textbf{DRF}         &           & 6.09e-06  &           & 1.37e-05*** \\
                     &           & (4.87e-06)&           & (5.10e-06) \\
\textbf{DPI}         &           & 0.00431   &           & 0.00348   \\
                     &           & (0.00704) &           & (0.00709) \\
\textbf{PD}          &           & -0.051    &           & -0.00971  \\
                     &           & (0.0652)  &           & (0.0412)  \\
\textbf{Constant}    & 178.2***  & 181.2***  & 214.2***  & 191.4***  \\
                     & (8.31)    & (11.73)   & (0.00830) & (0.00512) \\
\midrule
Observations         & 310       & 310       & 310       & 310       \\
R-squared            & 0.184     & 0.225     & 0.107     & 0.232     \\
\bottomrule
\end{tabular}
\begin{tablenotes}
\footnotesize
\item \textit{Note}: Standard errors in parentheses. \\
\item *** $p<0.01$, ** $p<0.05$, * $p<0.1$
\end{tablenotes}
\end{threeparttable}
\label{8}
\end{table}
\subsubsection{One-Period Lag Robustness Test}
Considering the possible time lag between the establishment of BRI node cities and observable improvements in provincial digital financial inclusion, a lagged treatment approach is employed, as in previous literature. All variables in the model are lagged by one period to test robustness. As shown in Model (4) of the robustness checks in \textbf{Table~\ref{8}}, if the policy takes effect one year after implementation, the impact on digital financial inclusion remains significant. This further confirms that the establishment of BRI node cities has had a substantial and robust effect on enhancing digital financial inclusion in the provinces.
\section{Main Findings and Policy Recommendations}

\subsection{Main Findings}

This study provides a detailed overview of the current state, policies, and development trajectory of China’s digital economy and digital inclusive finance. It also reviews the literature on the Belt and Road Initiative (BRI) and its impact on the digital economy and inclusive finance. Using panel data from 31 Chinese provinces covering the period from 2011 to 2020, the study applies a difference-in-differences (DID) approach to estimate the impact of the establishment of BRI node cities on provincial digital inclusive finance. The empirical results confirm that China’s digital economy policies significantly enhance digital financial inclusion. The main conclusions are as follows:

\begin{enumerate}
    \item The establishment of BRI and Digital Silk Road node cities significantly boosts the development of digital inclusive finance in the provinces where the node cities are located.
    \item Internet development, as a mediating variable, is positively influenced by the establishment of BRI node cities and in turn promotes the advancement of digital inclusive finance.
    \item The policy effects of BRI node cities on provincial digital inclusive finance exhibit regional heterogeneity, with central provinces benefiting more than western provinces.
\end{enumerate}

\subsection{Policy Recommendations}

Based on the empirical findings above, and to further advance the development of digital inclusive finance and the digital economy in BRI-participating provinces, this study offers the following policy recommendations across three dimensions—digital infrastructure construction, enterprise development, and digital trade—taking into account the developmental characteristics of different regions and industries:

\paragraph{First, digital infrastructure construction.} China should continue leveraging institutional advantages to accelerate digital infrastructure development in remote areas and improve urban-rural digital finance networks, thereby narrowing the digital divide. Compared with the United States, China's digital infrastructure is more regionally balanced, but still presents opportunities for improvement. Policies similar to the “Eastern Data, Western Computing” initiative should be further promoted to balance digital infrastructure development between eastern and western regions. Attention should also be paid to imbalances in digital infrastructure between highly urbanized and less urbanized areas. Agricultural provinces, especially major grain-producing ones, shoulder the national responsibility of securing arable land and food supply. Accelerating industrial digitalization in these provinces and enabling local technical staff and grassroots public officials to effectively utilize and maintain digital infrastructure are critical areas for future development.\cite{waldhauser2008large}

In this regard, the author believes that the establishment of BRI node cities represents an efficient solution. Through large-scale regional policies such as the Belt and Road Initiative and the Digital Silk Road, node provinces can be strategically selected and developed to enhance digital inclusive finance along the route. The regional spillover effects of these node cities and provinces can then more efficiently extend financial services across the region and even nationwide. By strengthening digital infrastructure, the “regional exclusivity” and “barrier effects” of traditional financial services can be mitigated, thereby improving the efficiency and quality of rural financial services and promoting digital agriculture and rural economic development.

\paragraph{Second, enterprise development.}  
To improve the efficiency of enterprise financing and reduce the burden of credit risk assessment for banks, China should vigorously promote the digital transformation of financial institutions and support the inclusive application of digital technologies in the financial sector. By leveraging technologies such as cloud storage, big data, the Internet of Things (IoT), mobile internet, artificial intelligence (AI), blockchain, and biometric recognition, financial institutions can optimize their service models and smoothly transition from traditional finance to digital finance.

From the perspective of enterprise digitalization, the government should provide subsidies to digital enterprises and encourage the cultivation of specialized talent to foster innovation in production models and expand the application scenarios of digital business. A user-centric development approach should be adopted—constructing user profiles and using big data networks to deliver more personalized and humanized digital services.

In terms of policy support, China has focused on improving the development environment for small and micro digital enterprises, which has laid a solid foundation for the digital economy and served as a key driver of innovation and research. Building on this foundation, further efforts should be made to improve the business climate for China’s cross-border digital enterprises in line with the Belt and Road Initiative and the Digital Silk Road, thereby promoting sustained growth in the digital enterprise sector.

\paragraph{Third, digital trade.}  
China has already achieved significant results through its promotion of the Digital Silk Road and Belt and Road Initiative. To consolidate these gains and further leverage the driving role of node cities in the development of digital inclusive finance, attention should be given to the following aspects:

First, China’s digital trade is facing challenges posed by aggressive U.S. trade policies. In response, China should deepen cooperation with countries along the Belt and Road route to build a more mature digital inclusive finance network. Joint development and mutual learning in technology can help counter U.S. trade barriers. Broader cooperation should also be established in areas such as cyberspace sovereignty and cross-border data transmission security, enhancing fairness in the international trade environment and breaking U.S. monopolies in various sectors of global digital trade.\cite{corrado2017inclusive}

Second, in deepening the BRI, China should use the initiative to promote the inclusive development of its domestic digital economy. This includes addressing internal shortcomings and driving technological progress in weaker digital sectors. The smooth operation of the domestic economic cycle is a crucial source of competitiveness for Chinese enterprises and products on the international stage.

Finally, a coordinated mechanism should be established between domestic digital economy development and international digital trade advancement to jointly promote a “dual circulation” development pattern that integrates domestic and international markets.

{\small
\bibliographystyle{bib}
\bibliography{main}
}

\clearpage

\end{document}